# Design of an Energy Aware peta-flops Class High Performance Cluster Based on Power Architecture

Wissam Abu Ahmad[1], Andrea Bartolini[2,3], Francesco Beneventi[2], Luca Benini[2,3], Andrea Borghesi[2], Marco Cicala[1], Privato Forestieri[1], Cosimo Gianfreda[1], Daniele Gregori[1], Antonio Libri[3], Filippo Spiga[4,5], Simone Tinti[1]

[1] E4 Computer Engineering, Scandiano (RE), Italy.
[2] DISI, DEI, University of Bologna, Bologna, Italy.
[3] Department of Information Technology and Electrical Engineering, ETH, Zurich, Switzerland.
[4] Quantum ESPRESSO Foundation, UK
[5] University of Cambridge, Cambridge, UK

wissam.abuahmad@e4company.com, a.bartolini@unibo.it, francesco.beneventi@unibo.it, luca.benini@unibo.it, andrea.borghesi@unibo.it, marco.cicala@e4company.com, tino.forestieri@e4company.com, cosimo.gianfreda@e4company.com, daniele.gregori@e4company.com, a.libri@iis.ee.ethz.ch, filippo.spiga@quantum-espresso.org, simone.tinti@e4company.com

*Abstract*—In this paper we present D.A.V.I.D.E. (Development for an Added Value Infrastructure Designed in Europe), an innovative and energy efficient High Performance Computing cluster designed by E4 Computer Engineering for PRACE (Partnership for Advanced Computing in Europe). D.A.V.I.D.E. is built using best-in-class components (IBM's POWER8-NVLink CPUs, NVIDIA TESLA P100 GPUs, Mellanox InfiniBand EDR 100 Gb/s networking) plus custom hardware and an innovative system middleware software.

D.A.V.I.D.E. features (i) a dedicated power monitor interface, built around the BeagleBone Black Board that allows high frequency sampling directly from the power backplane and scalable integration with the internal node telemetry and system level power management software; (ii) a custom-built chassis, based on OpenRack form factor, and liquid cooling that allows the system to be used in modern, energy efficient, datacenter; (iii) software components designed for enabling fine grain power monitoring, power management (i.e. power capping and energy aware job scheduling) and application power profiling, based on dedicated machine learning components. Software APIs are offered to developers and users to tune the computing node performance and power consumption around on the application requirements.

The first pilot system that we will deploy at the beginning of 2017, will demonstrate key HPC applications from different fields ported and optimized for this innovative platform.

## I. INTRODUCTION

It is nowadays broadly accepted that the trade-off between computing power and energy-efficiency is becoming of crucial importance for high performance computing evolution. Indeed, power consumption currently constrains the peak performance of large supercomputer installations and at the same time it is responsible for a significant slice of their TCO (Total Cost of Ownership). Accordingly to the Top500 list, which ranks supercomputers based on their peak performance (Flops - floating point operations per second - when running a Linpack benchmark) [1] [2], every new most powerful supercomputer has caused an increment of the total power consumption. This was true till Tianhe-2 (the former most powerful supercomputer, 1st from 06/2013 to 11/2015 Top500 lists), where the IT power consumption reached the practical limit of 17.8 MW for 33.8 PFlops.The current most powerful supercomputer TaihuLight reaches 93 PFlops with a power envelope of only 15.4 MW. This was possible thanks to an energy efficiency increment of 3x w.r.t. Tianhe-2. The fact that the improvement in energy efficiency matches the improvement in performance (33 to 93 PFlops) is a direct effect of hitting the power wall in supercomputing installations. We are clearly in an era of power-limited HPC evolution.

The Green500 list ranks Top500 supercomputers by an energy-efficiency metric, measured as Flops per Watt (Flops/W) [3]. From the Green500 perspective Sunway TaihuLight is ranked 4th with an energy-efficiency of 6 GFlops/W, while Tianhe-2 is ranked 135th with an energy-efficiency of 2 GFlops/W. Today the two most energy efficient supercomputers according to the Green500 list are DGX SaturnV and Piz Daint which achieve an energy efficiency of 9.5 and 7.5 GFlops/W respectively. In both cases this was possible thanks to a heterogeneous design based on the NVIDIA Tesla P100 [4] accelerator card, which is capable of up to 5.3 TFlops in double precision with 300 W of TDP.

However, the use of best-in-class hardware components alone cannot guarantee optimal energy-efficiency under supercomputer production workloads. Indeed, not all the production workloads have the same resource usage of the linpack, and an architecture optimized for it may fail in delivering the best trade-off between performance and energy efficiency when deployed. For this reason it is important to combine energy-efficient hardware with energy and power management strategies in a holistic design.

D.A.V.I.D.E. is the final stage of the European Prace Pre-

Commercial Procurement[1] (PCP) effort to develop an energy-aware High peta-flop class HPC machine.

The first sample nodes will be available from mid March 2017, and will be used for preliminary application porting and performance tuning. All the nodes will be assembled and tested using the E4 standard burn-in suite by the end of March. The whole system will be fully configured in April 2017 in the E4 facility in order to perform baseline performance, power and energy benchmarks using air cooling. It will be converted to liquid cooling starting from June 2017 then installed at CINECA premises. CINECA is a non-profit consortium of 70 Italian universities, the National Institute of Oceanography and Experimental Geophysics (OGS), the National Research Council (CNR), and the Ministry of Education, Universities and Reseach (MIUR). CINECA represents Italy in PRACE and is one of the four Tier-0 hosting centers.

The pilot system, that will be described in the following sections, is the result of a three-phases project. In the first two stages different technologies have been explored. Specifically the first two phases were based on multicore multiprocessor ARM 64-bit System On Chip due to the promising on the field test conducted on such platforms, including a previous prototype that lead to the design and manufacturing of an 80 TFlops ARM 64-bit + GPUs cluster [5], [6]. For the third phase ARM SoC have been replaced with IBM's POWER8-NVLink CPUs to exploit best-in-class acceleration technology which was not supported in ARM, as well as to exploit the mature software ecosystem as well.

In section II we will describe the hardware platform adopted to build the system, the optimized rack and cooling system based on commodity technology. In section III we will describe the additional hardware and software features which we have introduced for:

1) fine-grain monitoring of both the power and energy consumption;
2) managing power consumption by means of a proactive job-scheduler and power capper;
3) providing the programmer with APIs to adapt the node performance to the application characteristics.

In section IV we will describe our strategies for optimizing the set of application of European interest on the target architecture. This will involve application porting and tuning, looking at exploiting both POWER CPU and NVIDIA GPU strengths. The selected applications are the following:

- Electronic structure calculation and materials modeling at the nano scale with Quantum ESPRESSO [7] application.
- Oceanographic research and climatic studies with NEMO application [8].
- Geo-dynamics science with SPECFEM3D [9].
- Quantum Chromo Dynamics with Berlin QCD (BQCD) [10].

---

[1]Pre-Commercial Procurements are a way to obtain R&D servicies based on systems not yet available in the market. More informations available on line: http://www.prace-ri.eu/pcp.

In the conclusion we will summarize the results achieved during the development of this project and the roadmap of future developments.

## II. HARDWARE SETUP AND OPTIMIZATION

The hardware adopted in the D.A.V.I.D.E. supercomputer is based on commodity technology to be cost-effective but employs optimizations and advances in the cooling system and in the energy-efficiency support. Below the main elements that describe the system.

### A. POWER8 Processor Specifications

POWER8 architecture is designed to be a massively multithreaded CPU. Each of its cores is capable of handling eight hardware threads simultaneously, for a total of 96 threads executed simultaneously on an up to 12-core chip (D.A.V.I.D.E. supercomputer is powered by the 8 core version). The POWER8 core has 64 kB L1 data and 32 kB L1 instruction caches. Each core can issue ten instructions and can dispatch eight instructions per cycle to 16 execution pipes: two fixed-point pipelines, two load/store pipelines, two load pipelines, four double-precision floating-point pipelines that can also act as eight single-precision floating-point units, two VMX units, one cryptographic unit, one decimal floating-point units, one condition register unit, and one branch execution unit.

The memory module on the POWER8 chips are specified to use either DDR3 or DDR4 memory. Each memory module hosts a Centaur chip plus 16 MB of eDRAM that acts as a memory buffer (L4 cache). Every Centaur is connected to the POWER8 by means of three high-speed links (2:1 Read:Write), each running at 9.6 GB/s, for a total bandwidth of 28.8 GB/s, with a 40ns latency. Each POWER8 can be linked to up to eight Centaur chips allowing for up to 1 TB of memory per socket, with an aggregated 128 MByte L4 cache and 230 GB/s sustained memory bandwidth in and out of the processor.

D.A.V.I.D.E. supercomputer is based on an enhanced version of the CPU called POWER8+ that includes a new high speed bus (NVLink) used by the latest NVIDIA GPU based on Pascal Architecture.

### B. NVIDIA Pascal P100 NVLink GPGPU

The more the use of GPUs to accelerate compute applications has risen in recent years, the more the appetite for data in many of those applications has increased. Much larger problems are being solved by GPUs, requiring much larger datasets and higher demand for DRAM bandwidth. To address this demand for higher raw bandwidth, Tesla P100 is the first GPU accelerator to use High Bandwidth Memory 2 (HBM2), where the memory dies are linked using microscopic wires that are created with through-silicon vias.

NVLink uses NVIDIAs new High-Speed Signaling interconnect (NVHS). NVHS transmits data over a differential pair up to 20 Gb/s. Eight of these differential connections form a Sub-Link that sends data in one direction, and two sub-links, one for each direction form, a link that connects two processors (GPU-to-GPU or GPU-to-CPU). A single link supports up to 40 GB/s of bidirectional bandwidth between the endpoints. Multiple links can be combined to form "gangs" for even higher-bandwidth connectivity between processors. The NVLink implementation in Tesla P100 supports up to four links, enabling ganged configurations with aggregate maximum bidirectional bandwidth of 160 GB/s.

Tesla P100 was built to deliver high performance for the most demanding computing applications. Peak performance can be summarized as follows: 5.3 TFlops of double precision floating point (FP64); 10.6 TFlops of single precision (FP32); 21.2 TFlops of half-precision (FP16).

### C. OpenRack Liquid Cooled

Both the prototype phase II ARM cluster and the last stage peta-flops POWER cluster are equipped with direct hot-water cooling (35/40°C) for the CPUs and GPUs. We adopted a commercial solution made by Cool-IT Systems capable of extracting about 80% of the heat produced by the compute nodes, the remaining 20% will be dissipated using air cooling. This technology is extremely flexible and with minor modifications can fit in our infrastructure scheme. Each rack has an independent liquid-liquid or liquid-air heat exchanger unit with redundant pumps (see figure 1). The compute nodes are connected with the heat exchanger through pipes and a side bar for water distribution called manyfold. The minimal inlet temperature for the liquid that comes from the facility is 2°C, while the maximum is 45°C. The maximum outlet temperature of the facility liquid is 50/55°C. The temperature of the liquid that goes to the systems needs to be in the range between 5 degrees above dew point (to avoid any condensation on tubes, barbs and manifold ) and 45°C maximum temperature. Each CPU and GPU is cooled via a passive cold plate in direct contact with the silicon chip. This solution has been already used with success on Top500 machines with extremely good efficiency ratio.

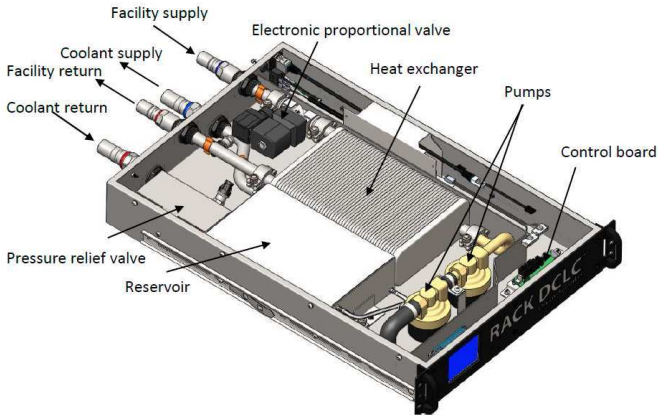

Fig. 1. Schematic representation of the liquid to liquid heat exchanger.

Having studied the Open Compute Project (OCP) and OpenRack standard (ORs) since Phase One, E4 decided to adopt these standards in the PCP proposal. This approach to the datacenter racks and nodes provides several benefits to power efficiency and serviceability, that will be detailed in section III.

### D. Integration Strategy

The technologies presented span over several domains of the computing facility and need to be correctly integrated. At node level several buses need to coexist. In particular:
1) the CPU uses a dedicated SMP interconnect to communicate,
2) the GPU transfers data throughput NVLink,
3) the high speed network card relies on PCI express Gen 3.

Inter-node communication is provided by an EDR Infiniband network. Each POWER8-NVLink CPU is connected to 2 NVIDIA Pascal P100 via PCIe to provide power and management communication, while the data movement takes advantage of the NVLink 1.0 bus with a bi-bandwidth of 80 GB/s, between both CPU-GPU and GPU-GPU. An external EDR infiniband network is plugged on to the 16x PCIe gen3 slot. To guarantee symmetric performance the configuration described above is replicated in its entirety for the second CPU of the same node. All compute nodes are connected via a dual 100 Gb/s EDR connection.

### E. Compute Node

The compute nodes derive from the OpenPOWER system designed with the codename Garrison. Each node will host two IBM POWER8+ with NVLink and 4 Pascal P100 with the intra node communication layout optimized for best performance. The original design of the server is air cooled, while the implementation for the Pilot system will use direct liquid cooling for CPUs and GPUs. A prototype version of the liquid cooling is shown in figure 2. Each compute node has a peak performance of 22 TFlops in double precision and an estimated power consumption of 2 kW.

### F. OpenRack form factor

Integration at rack level brings several advantages compared to the one at server level. From our analysis, the OpenRack or OCP standards provide a better TCO and higher reliability in large scale solutions. The consolidation of the power supply units (PSU) (reducing the total number from 2 per node to few per rack), removes several components with high failure rate (PSU have a high failure rate due to the stress on power on/off the node), and better distributes the power requirements during the boot procedure. In addition the reduction of the PSU number leads also to a reduction of up to 5% of the total power consumption due to a more efficient AC/DC conversion. Moreover, the quality of the power signal improves dramatically moving from AC/DC conversion at server level to AC/DC conversion at rack level. As reported in section III, thanks to that it is possible to have a low-noise measurement of the power consumption with high sampling speed (above 1kHz). A second important advantage of rack consolidation is the centralization of fans in the rear side of the rack, while server chassis are fanless. Finally, the rack integration guarantees a centralized redundant management module which simplifies all management procedures.

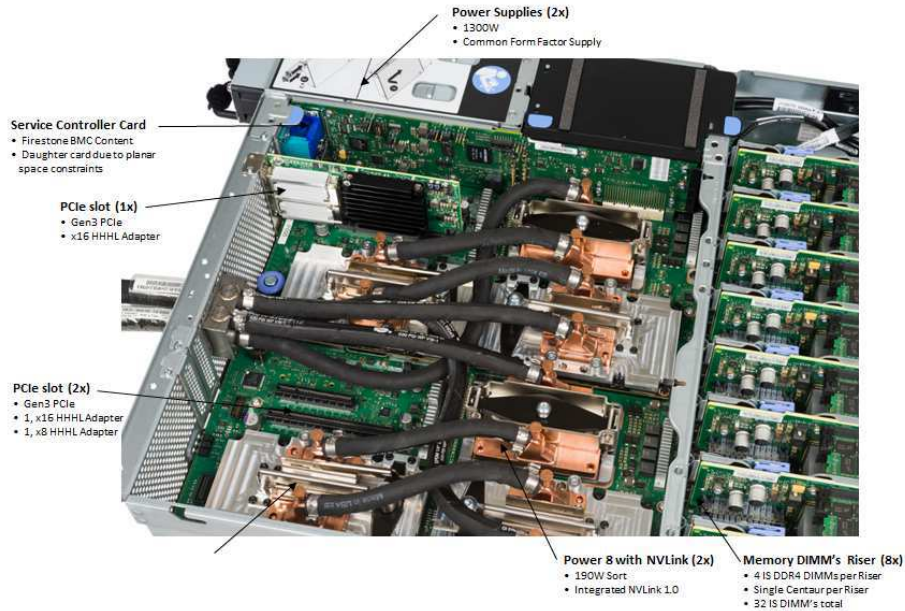

Fig. 2. Liquid cooled compute node.

*G. Cooling System*

The pilot system will adopt a variation of the Cool-IT solution. E4 has already tested this cooling approach in other installations with success, it is simple to install and it is certified against leakages. The depicted direct liquid cooling solution has been selected for two main factors:

1) an increase of the energy-efficiency of the system;
2) and to ensure a safe operating temperature in the silicon die during job execution.

All power hungry components (CPUs, GPUs, DIMMs) are throttled when a maximum operating temperature is reached. This often happens in air cooled server, causing an overall performance degradation, which is normally not evenly distributed across the server nodes. Direct liquid cooling solves this issue by providing to all the CPUs and GPUs components the same cooling capacity. D.A.V.I.D.E. will remove 75-80% of the heat through direct liquid cooling, the remaining 20-25% is removed with a heavy duty low speed fans. As shown in figure 3, the fans are much larger compared to the standard server one.

*H. Network*

D.A.V.I.D.E. will feature a high speed network EDR infiniband with one card per CPU socket. We will use a dual plane configuration on the top system to avoid comminication on the SMP. The aggregate bandwidth per node is 200 Gb/s. The topology will be fat-tree with no oversubscription.

*I. Self-contained pilot system*

The Pilot system is designed to be completely self consistent. Which means that it provides all components integrated in a whole HPC solution. Each rack needs to be connected to the inlet and outlet water connectors and to a power line capable to provide 32 kW. To connect the system there are 2x10 Gb/s Ethernet uplinks SFP+. The system consists of four racks, 3 of them dedicated to the compute nodes and 1 to the storage, management and login nodes. Each rack has a weight of 800 kg and dimension of 800 mm width x 1200 mm depth x 2500 mm high. The estimated total power consumption of the Pilot system is less than 100 kW with a peak performance of 1 PFlops. The flow rate estimated is 30 L/min per rack with water temperature at 35°C. It is estimated a dissipation between 75% to 80% of the heat through direct liquid cooling.

## III. HARWDARE AND SOFTWARE ENERGY AWARE CUSTOMIZATION

In D.A.V.I.D.E. the standard cooling system has been modified in order to work with 21 inch width server in a OpenRack infrastructure.

Unlike the traditional approach where each system fitted into the rack is a full featured server, with the OpenRack the redundant functions between all the different rack-mount systems are removed and consolidated into the rack itself. This leaves primary rack-mounts that serves a main function only, such as pure computing, storage, or computing/storage hybrids, and support modules providing the previously redundant functions. Such support modules, all located at the rear of the rack, include:

1) Fan modules that accommodate heavy duty 5U fans, much bigger than the ones usually installed in standard

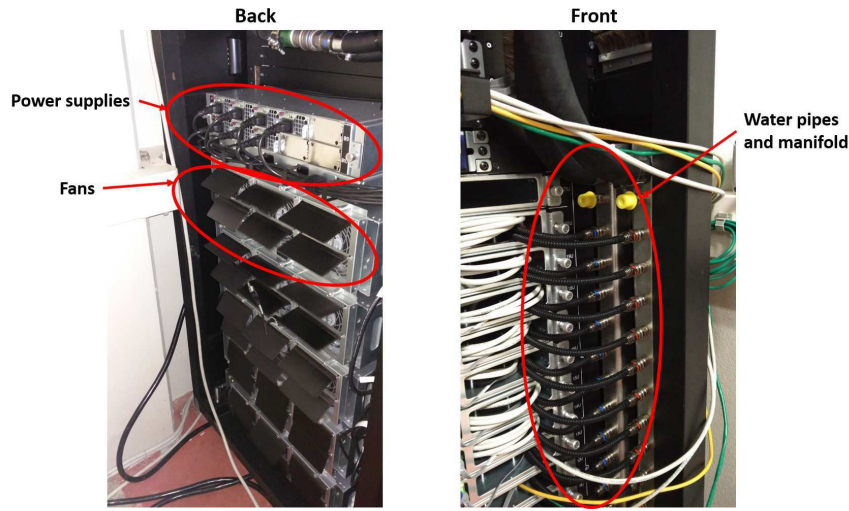

Fig. 3. OpenRack details: upper-left, Fans and PSU. Upper-right, liquid cooling manifold. Lower, redundant management control.

rack-mounts. They deliver significantly better cooling performance and power efficiency.

2) A power bank module that can receive enough standard server power supply units for the whole rack's needs (supports up to 32 kW). It regroups and connects them to a rack-level power distribution unit in which the primary rack-mounts can be hot-plugged, in the same fashion as a storage backplane. This power distribution unit and the primary rack-mounts connectors are all made of pure copper for the best power distribution efficiency. This consolidation of the power supplies in the rack provides redundancy to all the servers installed in to the rack.

3) A remote management controller module, serving as a gateway for the management related traffic between the sub-rack and super-rack levels. This module is capable, among others, of real time fan speed optimization, comprehensive rack asset management (with rack IDs, node IDs, asset tags, and so on), and full featured power management.

### A. Design for energy efficiency

D.A.V.I.D.E. is designed to support energy and power aware usage. Figure 4 depicts its main functionalities.

Each node of D.A.V.I.D.E. contains a dedicated programmable SoC that serves as an energy and power gateway. The energy-gateway (EG) is connected to power sensors. This allows measurement with high sampling rate the power consumption of the entire node, as well as of the main computing components in the node. The energy-gateway supports network time protocols. Thanks to that, D.A.V.I.D.E. provides synchronous time-stamping for the power and energy measurements that can be correlated within the nodes as well as with profiling information coming from the computing resources. In addition, the energy-gateway in D.A.V.I.D.E. supports a machine-to-machine (M2M) protocol, this allows flexible integration of the power and energy information with the system software components. At the higher level, the job scheduler features a dedicated plugin to receive the monitoring information and to correlate them with user requests and scheduling decisions. This correlation enables per user and per job energy-accounting (EA) and profiling (Pr). In parallel, this information is recorded into a database, and computed by the management node for the training of job-to-power predictors (EP) based on the historical job request and power traces. Once available, the trained power predictors are used by the job scheduler to constrain the total power consumption of the D.A.V.I.D.E. supercomputing machine. The power cap can be specified by the system administrator to follow infrastructure requirements. For this goal, the job scheduler and dispatcher is augmented by a dedicated engine. This is capable of using a per job power prediction to select which job should enter the supercomputing machine at each moment, in order to fulfill the specified power envelope while preserving job fairness. At the middle layer, each D.A.V.I.D.E. node features a dedicated energy-proportionality support and APIs that enable each node to activate specific computing resources, or limit their performance and features accordingly to the job requirement and request. As an effect of this, each computing node can be tailored to the job requirement, achieving a deeper energy-efficiency.

*1) Energy and power gateway:* At a higher level, power information is used by the job scheduler to both (i) to account for the energy-to-solution (ETS) of each job/user, and (ii) control the overall power consumption of the supercomputer. The former allows the energy consumption cost of each job to be distributed between the supercomputng center and the user, promoting an energy-aware usage of the resources. The latter, instead, aims to reduce the operating costs.

At a lower level, fine grained power measurements are used by both the energy-aware run-time and the power/thermal management software to dynamically improve the energy-efficiency, imposing power and thermal constraints. In this context, power measurements need to be used within the node by dedicated control-agents, and, at the same time, by centralized per-job aggregator agents. For this reason, the measured values need to be available in real-time to multiple agents with a low-latency and a synchronized timestamp.

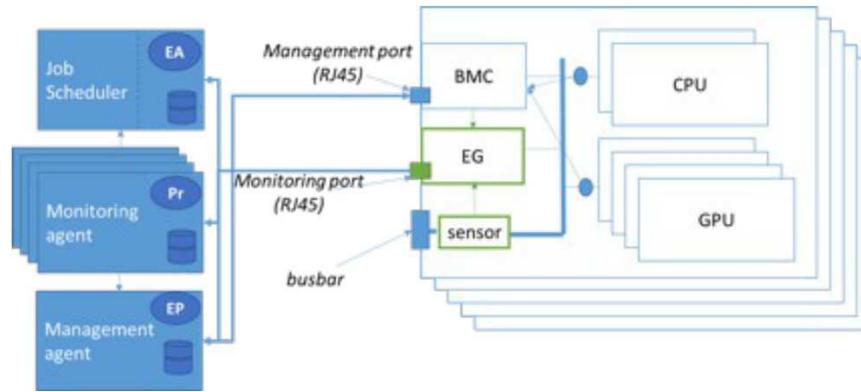

Fig. 4. Main functionalities of D.A.V.I.D.E. power aware usage.

Finally, at user level the power measurements are needed by profiling tools, to correlate the power consumption with program phases and architectural events, and by accounting tools to implement accounting strategies, based on the actual user energy consumption. In this context, power measurements need to be synchronized with the application phases without introducing performance loss in the computing nodes during the monitoring.

In light of the above, an ideal node power monitoring system needs to (i) support an accurate and fast sampling, (ii) support a scalable sharing of the power values to multiple agents, (iii) be external from the computing resources of the node, and finally (iv) be synchronized with them.

In D.A.V.I.D.E. all of these requirements are fulfilled thanks to an additional component, namely the energy and power gateway (EG) which combines an embedded SoCs with a scalable, standard and open communication protocol, namely the *Message Queuing Telemetry Transport (MQTT)* protocol [11], which organizes the data-exchange in a topic/subscriber approach. The EG can be easily re-programmed to build on top of the MQTT communication emerging power measurement APIs (e.g. PowerAPI [12]), aiming to standardize the power measurement interface. We exploit the built-in Analog-to-Digital Converter (ADC) in the embedded SoC, and its advanced functionalities, to provide a fine-grain and accurate sampling up to 800 kS/s, decimated to 50 kS/s, on power-measuring sensor outputs. The EG is powered by a Beaglebone Black Board (BBB) originally designed for IoT applications. It includes the *TI Sitara AM335x* SoC, which is based on an ARM®Cortex-A8 processor. It has a built-in 12-bit SAR ADC supporting up to 1.6 MS/s, with 8 multiplexed channels, and integrates hardware-support for device synchronization via the Precision Time Protocol (PTP) [13].

The BBB also serves as an hub for the all node telemetry. In the D.A.V.I.D.E. computing nodes, not only node power is accessible at high accuracy, but also both per component power consumption and architectural events can be monitored out-of-band from the BBB, and sent to external agents and smart profilers (Pr). Indeed, such a monitoring runs data intelligence on the monitored data to identify sources of not-optimality and hazards.

*2) Power-Aware System Management:* Limiting power and energy consumption is a growing issue in the HPC world and the subject of numerous research works, targeted at either developing new hardware and software solutions or optimizing the management of existing systems. Many strategies try to limit the power consumption within a certain power budget, never to be exceeded. These methods are generally referred to as *power capping* [14]. A big challenge for the adoption of power capping solutions is the need to find a good balance between curtailing the power consumption and keeping a high level of Quality-Of-Service for the system users. D.A.V.I.D.E. computing nodes are capable of controlling in real-time the power consumption of the node and of its internal components. When activated, a total node power cap is maintained by local feedback controllers which tune the operating points of the internal components in the compute node to track the maximum power set point. When used as the only mechanism to manage the total system power consumption, node level power capping can lead to performance loss and Service Level Agreement (SLA) violation.

To overcome this limitation, a key role can be played by the scheduling software that decides where and when a job has to execute, a software module commonly referred to as *job dispatcher*. With a "clever" job dispatcher it is possible to operate a power capped system at a high Quality-of-Service: the main idea is to act on the job execution order alone. Previous works shown that extending current HPC system job dispatchers with power capping could lead to substantial energy savings without degrading the performance of the supercomputer and the QoS for the users [15] [16]. The approaches studied in these works produce *proactive* schedules: they consider all the jobs which need to be run - and submitted in a job queue by the supercomputer users - and decide the starting time of each of them in advance, according to a specified objective

(i.e. QoS, maximal power savings, etc.). Works in [17] [18] show that job power consumption can be estimated before job execution, based on user's request and at job submission information. D.A.V.I.D.E. will support the creation of per-job power estimators and will take advantage of their predictions in the job scheduler.

The D.A.V.I.D.E. management system aims to mix both proactive and reactive power capping techniques, striving to combine the benefits from both worlds. To this end, we are going to extend the basic functionality of D.A.V.I.D.E. resource manager, which is SLURM (Simple Linux Utility Resource Management) [19]. SLURM is an open-source cluster resource management system designed to be flexible and fault-tolerant. Its flexibility allows its behavior to be changed according to our goal, and to implement novel management policies through the development of new software plug-ins. We plan to act on the job dispatching component, i.e. modifying the order in which jobs are executed and the resource selection process, and add frequency scaling/socket-level power cap capabilities in order to reduce the power consumption of the whole system.

## IV. APPLICATION PORTING AND TUNING

In the race toward Exascale, lot of emphasis is set on the concept of *co-design* to bring together, starting from the early design stage of a HPC system, hardware, software and middleware (libraries, compilers, run-times) in order to achieve the maximum usability and performance. During the co-design phase of the GPU-accelerated system D.A.V.I.D.E., four applications of European interest have been considered: Quantum ESPRESSO, NEMO, SPECFEM3D, Berlin QCD (or BQCD).

Looking at the literature and the HPC prototypes developed and deployed in Europe, co-design for energy efficiency mostly involves only hardware components. We believe co-design for energy efficiency can only be effective if it also involves the software. A typical scenario is a process in the background (standalone or driven by the resource manager) which detect the utilization of various system components and then autonomously decides, based on predefined heuristic, to throttle or power down one component (for example unused cores). In order to make the co-design for energy efficiency process application-driven, we believe it is important to empower the application developer with the ability to feed back to the system useful information that can have an effect on the energy to solution of each application run. Looking at today and future NVIDIA GPU accelerators, the amount of compute power per single accelerator will increase dramatically compared to CPUs with higher efficiency. A general strategy for energy efficiency in a GPU system is to leverage as much as possible the accelerator by enabling much of the application wall time to run on GPU, and avoiding too frequent data transfers from GPU memory to CPU memory (ideally limiting them only when MPI communication is needed).

In collaboration with the ETH Multitherman Laboratory we are designing a set of APIs to switch off or put in sleep mode particular system components on-demand, such as unused CPU cores, memory controllers and GPU. These APIs will be wrapped in the job scheduler to size the node around the job requirements as well as around a library that application developers will explicitly call inside the source code and link during compilation. By empowering application developers to consider energy implication during the development phase, complex codebase coarse-grain regions for energy saving opportunities can be identified. Only application developers have the appropriate knowledge to understand the code and visualize almost immediately the impact of these changes applied to multiple scenarios without necessary run a huge time-consuming set of benchmarks. Once an application is instrumented, the system will react and measure based on API calls. From a co-design point of view, by iterating multiple times coding and experiments, application developers can compare time-to-solution versus energy-to-solution and identify the right tradeoff between each application.

### A. Quantum ESPRESSO

Quantum ESPRESSO is an integrated suite of computer codes for electronic-structure calculations and materials modeling, based on density-functional theory, plane waves, and pseudo-potentials (norm-conserving, ultra-soft, and projector-augmented wave). Quantum ESPRESSO is written mostly in Fortran 90 with few portions in C. It uses MPI for multi-node parallelization and it supports OpenMP for intra-node parallelization. It relies on multiple external libraries, including BLAS and LAPACK and FFT. Quantum ESPRESSO now supports GPU via a GPU plugin written in C and interfaced to Fortran code. A new improved code written in CUDA FORTRAN is going to be available by the time D.A.V.I.D.E. will be deployed.

Quantum ESPRESSO bottlenecks are various and different input cases demonstrate to stress different part of the code in different ways. It is well understood that one of the major performance impact factors is in the Fast Fourier Transform (FFT), especially at nanoscale. In order to exploit high memory bandwidth in Pascal GPU, the FFT computation has to be localized on the GPU memory cutting off as much MPI communication as possible. Because of the presence of NVLink, peer-to-peer GPU-to-GPU communication, allowing to localize FFT computation in group of 2 GPUs, avoids intra-node communication.

### B. NEMO

NEMO (Nucleus for European Modelling of the Ocean) is a state-of-the-art modelling framework for oceanographic research, operational oceanography seasonal forecast and climate studies. It consists of modular "engines" which can be combined as configurations to study interactions with ocean currents, sea-ice and other components of the climate system, such as vegetation and land surfaces. NEMO is utilized by a large international community, and its evolution is controlled by an European Consortium between multiple European countries: CNRS (France), Mercator-Ocean (France), NERC (UK), UKMO (UK), CMCC (Italy) and INGV (Italy). NEMO is mostly written in Fortran 90, designed to be highly-portable and parallelised using MPI with a regular domain decomposition in latitude/longitude. NEMO is essentially a stencil-based code with limited parallelism, low computational intensity and frequent halo exchanges.

Accelerating NEMO for GPU has been an effort attempted by multiple institutions in multiple occasions, but a GPU implementation is not yet converged into the official source code. Because of its flat timing profile (in early benchmarks, not a single routine consume more than 15% - 20% of the runtime), OpenACC [20] is the perfect choice: minimal changes on the source code; the final code runs on CPU if compiled without OpenACC; more maintainable and quicker to adapt to changes. Early experiments show dominance of memory-bandwidth bound kernels on GPU. An old OpenACC implementation will be ported to a recent version of NEMO which has an improved MPI implementation and better scalability. NEMO allocates a huge amount of data structure during its life time, and availability of memory on the GPU can become the bottleneck for very big input cases. Because of NVLink and the high memory bandwidth of the POWER system, NEMO will going to be a good test case to evaluate the quality and the driver runtime implementation of NVIDIA Unified Memory.

### C. SPECFEM3D

SPECFEM3D simulates three-dimensional global and regional seismic wave propagation based upon the spectral-element method (SEM). Two versions of the package are specifically designed to simulate seismic wave propagation at different scales: the "Cartesian" version for local simulations and the "Global" version for the simulations at the scale of the globe. We will focus on "Global" for our future investigation. SPECFEM3D Globe is written in Fortran2003 and other part in C, CUDA and OpenCL [21]. It is a highly scalable code using non-blocking MPI communications and includes several performance improvements in mesher and solver. In the official source code there is GPU support for both OpenCL and CUDA hardware accelerators, based on an automatic source-to-source transformation technique.

SPECFEM3D (both Globe and Cartesian) computational kernels should definitely benefit from the increased bandwidth of Pascal GPU versus previous generation, as well as the larger memory capacity. For parallel GPU scaling, it basically just does boundary exchanges which are all already neatly overlapped. The overall performance is not affected by messaging passing overhead as long as you have sufficient amount of work per GPU. In term of NVLink benefits, an increment of performance by explicitly handled GPU-to-GPU communications can be achieved for problem sizes which fit a single node. This subset of problems occur mostly for SPECFEM3D Cartesian for Oil&Gas simulation, rather than for SPECFEM3D Globe.

### D. Berlin QCD

BQCD (Berlin Quantum ChromoDynamics program) is a Hybrid Monte-Carlo code that simulates Lattice Quantum Chromodynamics (LQCD) with dynamical Wilson-type fermions. The computations take place on a four-dimensional regular grid with periodic boundary conditions. The main kernel of BQCD is a conjugate gradient solver with even/odd preconditioning. Within this kernel, a matrix-vector multiplication, where the matrix is sparse, is the dominating operation. Parallelization is based on regular domain decomposition in up to 3 dimensions. Most parts of the code are written in Fortran 90. It is parallelized using MPI and OpenMP. It relies on few external libraries for data portability across lattice QCD packages (e.g. LIME - Lattice QCD Interchange Message Encapsulation). The computational core of BQCD is the QUDA [22] [23] [24] library, which handle the specialized conjugate gradient for QCD calculations.

The latest version of QUDA supports construction of the clover field in QUDA. This avoids needing to compute it on the CPU, which can be costly compared to the subsequent GPU-accelerated Conjugate Gradient linear solver, and avoids the PCIe transfer overhead of copying it from the CPU. All auxiliary force computations are supported in QUDA, so that an entire Hybrid Monte-Carlo gauge generation can be offloaded to GPUs. This part is currently missing in the GPU-accelerated version of BQCD available on the web, the porting activity will enable the latest QUDA on a recent version of BQCD to support all these new features and performance improvements. From a computational library perspective, recent releases of QUDA have continued to improve the strong scaling on GPUs, including direct peer-to-peer communication between GPUs in the same node, removing MPI overhead. This drastically improves the scaling within dense GPU systems, making QUDA leveraging NVLink transparently and scaling within dense nodes nearly perfect.

## V. RELATED WORKS

### A. Heterogeneous Computing

Heterogeneous computing is an effective strategy to improve energy-efficiency, it consists of coupling general purpose processors with data parallel or massively parallel computing engines. This is visible in the Green500 list [1], where in the top ten most energy-efficient supercomputers, 9 out of 10 use accelerators while in the top ten positions of the Top500 list [3], only 5 out of 10 use an heterogeneous design. In addition, NVIDIA Pascal P100 as used in D.A.V.I.D.E. are present in the two most energy-efficient computing systems.

### B. Liquid Cooling

In a supercomputer a significant portion of the power consumption is spent in removing heat produced by the computation. Traditional cooling methods, based on Computer Room Air Conditioners (CRAC), or Computer Room Air Handlers (CRAH) (depending if the evaporator is built into the AC unit or an intermediate coolant is used [35]), have been enhanced with a free cooling mode, i.e. the capability to exploit the outside air, using only the AC blowers to circulate it in the room [36]. In addition, liquid-based technology is frequently combined with classical air-based systems, in the so-called hybrid cooling solutions, in order to take advantage of the superior heat removal capacity of liquid coolant (commonly water) with respect to air [37]. Liquid can be either conducted to heat exchangers directly connected to the most thermally critical IT equipment (usually the processing elements) [38], or made to flow through liquid-to-air Heat Exchangers. Moskovsky et al. analyse the implication of hot water cooling in a real supercomputer installation. By using a liquid at higher temperature to refrigerate the computing resources, it is possible to extend effectiveness of free-cooling [39]. However, IT performance degrades with the increased water temperature. D.A.V.I.D.E. uses a hybrid cooling approach and supports hot-water cooling.

### C. Energy-Power monitoring

Hackenberg et al. [25] highlight the importance of several factors in the power monitoring system, such as the accuracy of the power sensors and their acquisition chain, but also the sampling frequency and the timestamp synchronization, when used to accurately measure the energy consumption and correlate it with workload traces among several nodes. Today a per-node power profile can be obtained from the node BMC through the IPMI interface. However, this mechanism is characterized by a slow sample-rate (in the order of seconds) without time-stamping, and is affected by aliasing noise. The reason for the latter is that the power is measured instantaneously and cannot be used to account for the energy consumed in-between two subsequent power measurements [25] [26].

To overcome these problems, authors of [25] [26] propose the High Definition Energy Efficiency Monitoring (HDEEM) infrastructure. This system extends the BMC power/energy measurement features by inserting a set of Hall-Effect sensors in series on the different power-lines, which are then routed to the BMC by means of an FPGA. This framework is capable of fast sampling (up to 8 kS/s) the power consumption without aliasing and, thanks to its accurate time-stamping, allows the correlation of workload and power-traces, as well as accurate energy monitoring. However, all the power measurements in HDEEM are gathered through the BMC. As the BMC provides system management control, it is usually accessible only by system administrators and has a closed interface which cannot be easily reconfigured. In order to tackle closed designs and configuration interfaces, few attempts have been made to use open-source and low-cost embedded-computers. ArduPower [27] is based on Arduino Mega 2560 and is used to read the power-sensors output, while PowerInsight [28] uses a BeagleBone Black to read the power sensors through external ADCs. While both these design use an open-source System-on-Chip (SoC), they only provide measurements through custom interfaces which cannot be easily integrated with other system components. In addition, they reach only up to 1 kS/s, due to the use of external ADCs and a non-optimized software stack running on the embedded SoC.

In D.A.V.I.D.E. the energy gateway is based on a completely open design based on a low-cost and open-source state-of-the-art IoT board, and on an open implementation of the MQTT protocol. This enhances the performance of previous solutions by supporting the measurement of the power consumption at 800 kS/s averaged in HW to 50 kS/s.

### D. Job Scheduler

As discussed in subsection III-A2, limiting power and energy consumption has been a growing issue in the HPC world and the subject of numerous research works.

To overcome this problem, many approaches rely on decreasing the performance of the computing nodes, which in turn leads to increased duration of HPC jobs or longer waiting times. The two most common methods to achieve this are 1) *Dynamic Voltage and Frequency Scaling* (DVFS) [29] and 2) Intels *Running Average Power Limit* RAPL [30], [31]. These two techniques can be defined as *reactive*, meaning that they dynamically react to the system conditions and workload, and adjust the power through hardware-based mechanisms at runtime.

With DVFS, a processor can run at one of the supported frequency/voltage pairs lower than the nominal one. The main issue with DVFS-based approaches is the trade-off between power savings and decrease in performance: reducing the operating clock increases the duration of the applications which run on the slowed-down resources. To overcome this issue, several methods try to apply DVFS only in periods of low system activities [32] or in particular phases of a job execution [33]. While DVFS works by selecting a frequency or voltage among the range of possible ones, RAPL is a management interface that only requires the user to define a power threshold. The internal hardware then performs automatic frequency scaling and power throttling in order to keep the power consumption within the user-specified limit. RAPL employs an internal model of energy consumption to compute the average power consumption over a time frame, and tries to enforce the power cap as precisely as possible. The main advantages of RAPL w.r.t. DVFS are the integration of power monitoring and control inside the chip and a finer granularity (compared to the finite set of frequencies of DVFS). An important aspect of RAPL-bases technique is the decision of the amount of power to allocate to each computing node: for example, algorithms that aim at sharing the available power among the nodes can lead to good results in terms of QoS [34].

D.A.V.I.D.E. extends the job scheduler with a machine learning engine to train job power predictors, and with dedicated proactive and reactive energy-aware job scheduling policies.

## VI. CONCLUSION

In this paper we described the D.A.V.I.D.E. supercomputer based on commodity hardware and some hardware and software customization to improve energy efficiency.

With our prototype we demonstrate that it is possible to integrate Cost Effective Technologies to achieve high performance and significatively improve energy efficiency, thanks to the development of custom hardware and software components manageable with APIs.

This system is the building block for the forthcoming exascale supercomputer based on a class of system where Energy Aware management is mandatory.

## VII. ACKNOWLEDGMENT

## REFERENCES

[1] *Top500 List*, Available on line at: https://www.top500.org/
[2] *HPL - High-Performance Linpack Benchmark*, Available on line at: http://www.netlib.org.benchmark/hpl/
[3] *Green500 List*, Available on line at: https://www.top500.org/green500/
[4] *P100 - Gp100 Pascal whitepaper*, Available on line at: https://images.nvidia.com/content/pdf/tesla/whitepaper/pascal-architecture-whitepaper.pdf
[5] Nikola Rajovic et al., *The Mont-Blanc prototype: an alternative approach for HPC systems*. Proceedings of the International Conference for High Performance Computing, Networking, Storage and Analysis, Salt Lake City, Utah, 2016. ISBN: 978-1-4673-8815-3.
[6] D. Bortolotti et al., *User-space APIs for dynamic power management in many-core ARMv8 computing nodes*. International Conference on Hihg Performance Computing & Simulation (HPCS), 2016.
[7] *Quantum-Espresso*. Available Online at: http://www.quantum-espresso.org/
[8] *NEMO*. Available Online at: http://www.nemo-ocean.eu/
[9] *Specfem3D*. Available Online at: https://geodynamics.org/cig/software/specfem3d/
[10] Yoshifumi Nakamura and Hinnerk Stuben, *BQCD Berlin quantum chromodynamics program*. PoS Lattice2010:040,2010. Available Online at https://arxiv.org/abs/1011.0199
[11] *MQTT*. Available Online at: http://mqtt.org/
[12] R.E.Grant et al., *Standardizing Power Monitoring and Control at Exascale*, Computer Vol.49 Oct.2016, ISSN: 0018-9162
[13] A.Libri et al.: *Evaluation of synchronization protocols for fine-grain HPC sensor data time-stamping and collection*. 2016 International Conference on High Performance Computing & Simulation (HPCS), Innsbruck, 2016, pp. 818-825. doi: 10.1109/HPCSim.2016.7568419
[14] C. Lefurgy et al.: *Power capping: A prelude to power shifting*. Comput (2008) 11:183. doi:10.1007/s10586-007-0045-4.
[15] A.Borghesi et al., *MS3: A Mediterranenan-stile job scheduler for supercomputers - do less when it's too hot!*. International Conference on High Performance Computing & Simulation (HPCS), 2015.
[16] A.Borghesi et al., *Power Capping in High Performance Computing Systems*. Lecture Notes in Computer Science, Vol 9255, ISSN 0302-9743. 2015
[17] A.Borghesi et al., *Predictive Modeling for Job Power Consumption in HPC Systems*. High Performance Computing: 31st International Conference, ISC High Performance 2016, Frankfurt, Germany. ISBN: 978-3-319-41321-1.
[18] A.Sîrbu et al, *Power Consumption Modeling and Prediction in a Hybrid CPU-GPU-MIC Supercomputer*. Springer 2016. ISBN:978-3-319-43659-3.
[19] M.A.Jette et al., *SLURM: Simple Linux Utility for Resource Management*. In Lecture Notes in Computer Science: Proceedings of Job Scheduling Strategies for Parallel Processing. Springer-Verlag 2003
[20] *OpenACC*, Available OnLine at: http://www.openacc.org
[21] *OpenCL*, Available OnLine at: https://www.khronos.org/opencl/
[22] *QUDA*, Available OnLine at: https://lattice.github.io/quda/
[23] M. A. Clark et al., *Solving Lattice QCD systems of equations using mixed precision solvers on GPUs*. Comput. Phys. Commun. 181, 1517 (2010).
[24] R. Babich et al., *Scaling lattice QCD beyond 100 GPUs*. International Conference for High Performance Computing, Networking, Storage and Analysis (SC), 2011.
[25] D. Hackenberg et al., *HDEEM: High Definition Efficiency Monitoring*. Energy Efficient Supercomputing Workshop (E2SC), 2014.
[26] T.Ilsche et al., *Power Measurements for compute nodes: Improving sampling rates, granularity and accuracy*. Green Computing Conference and Sustainable Computing Conference (IGSC), 2015.
[27] M. F. Dolz et al., *ARDUPOWER: A low-cost wattmeter to improve energy efficiency of HPC applications*. Green Computing Conference and Sustainable Computing Conference (IGSC), 2015.
[28] J. H. Laros et al., *PowerInsight - A commodity power measureement capability*. International Green Computing Conference (IGCC), 2013.
[29] Hsu Chung-hsing et al., *A Power-Aware Run-Time System for High-Performance Computing*. Proceedings of the 2005 ACM/IEEE Conference on Supercomputing. ISBN: 1-59593-016-2.
[30] Intel Corporation *Intel® 64 and IA-32 Architectures Software Developer's Manual*. 2009
[31] D. Howard et al., *RAPL: Memory Power Estimation and Capping*. Proceedings of the 16th ACM/IEEE International Symposium on Low Power Electronics and Design, 2010. ISBN: 978-1-4503-0146-6.
[32] E. Maja et al., *Utilization driven power-aware parallel job scheduling*. Computer Science - Research and Development, 2010, Vol. 25, nr. 3. ISSN:1865-2042.
[33] B. Rountree et al., *Adagio: Making DVS Practical for Complex HPC Applications*. Proceedings of the 23rd International Conference on Supercomputing 2009. ISBN: 978-1-60558-498-0.
[34] D.A. Ellsworth et al., *Dynamic Power Sharing for Higher Job Throughput*. Prooceding of the International Conference for High Performance Computing, Networking, Storage and Analysis, 20015. ISBN: 978-1-4503-3723-6.
[35] T. Evans, *The different types of air conditioning equipment for IT environments*. APC White Paper n. 59, pp. 119, 2004.
[36] M. Pore et al., *Techniques to achieve energy proportionality in data centers: A survey*. in Handbook of Data Centers S. U. Khan and A. Y. Zomaya, Eds. Berlin: Springer, 2015.
[37] D. F. M. K. Patton, *The state of data center cooling: A review of current air and liquid cooling solutions*. Intel White Paper, pp. 111, 2008.
[38] L. Li, X. W. W. Zheng, and X. Wang, *Coordinating liquid and free air cooling with workload allocation for data center power minimization*. Proc. of ICAC, pp. 249259, 2014.
[39] A. Moskovsky et al., *Server level liquid cooling: Do higher system temperatures improve energy efficiency?*. Supercomputing frontiers and innovations, 3(1):67-74, 2016.